\documentclass[runningheads]{llncs}
\usepackage{graphicx}
\usepackage{todonotes}
\begin{document}
\title{Older Adults' Motivation and Engagement with Diverse Crowdsourcing Citizen Science Tasks}
\titlerunning{Older Adults' Engagement with Crowdsourcing Citizen Science Tasks}

\author{Kinga Skorupska\inst{1}\orcidID{0000-0002-9005-0348} \and
Anna Jaskulska\inst{2}\orcidID{0000-0002-2539-3934} \and
Rafał Masłyk\inst{1}\orcidID{0000-0003-1180-2159} \and
Julia Paluch\inst{1}\orcidID{0000-0002-7657-7856} \and
Radosław Nielek\inst{1}\orcidID{0000-0002-5794-7532} \and
Wiesław Kopeć\inst{1}\orcidID{0000-0001-9132-4171}}
\authorrunning{K. Skorupska et al.}
\institute{Polish-Japanese Acaemy of Information Technology, Warsaw, Poland 
\email{kinga.skorupska@pja.edu.pl}\\
\and
Kobo Association, Warsaw, Poland
}

\maketitle              % typeset the header of the contribution
\begin{abstract}
 In this exploratory study we evaluated the engagement, performance and preferences of older adults who interacted with different citizen science tasks. Out of 40 projects recently active on the Zooniverse platform we selected top ones to be represented by 8 diverse, yet standardized, microtasks, 2 in each category of image, audio, text and pattern recognition. Next, 33 older adults performed these microtasks at home and evaluated each task right after its completion to, finally, share what could encourage them to engage with such tasks in their free time. Based on the results we draw preliminary conclusions regarding older adults' motivations for engaging with such crowdsourcing tasks and suggest some guidelines for task design while discussing interesting avenues for further inquiry in the area of crowdsourcing for older adults.

\keywords{Crowdsourcing \and Older adults \and Citizen science \and Motivation.}
\end{abstract}

\section{Introduction and Related Works}

The area of crowdsourcing for older adults is both underappreciated and underexplored and developing sustainable solutions for older adults is still challenging \cite{knowles2018wisdom,knowles2018older}.
This may be due to multiple barriers both specific to the required ICT-skills \cite{aula_learning_2004} and the nature of crowdsourcing microtasks. Older adults differ from the younger generation in their online behavior and decision-making \cite{von2018influence} and they seem more selective when choosing their engagements \cite{djoub_ict_2013}, which, alongside their generally lower ICT skills, may explain how little interest they expressed in the Mechanical Turk platform populated by tedious and repetitive tasks \cite{brewer2016would} and lacking a suitable motivation to participate in crowdsourcing, as tasks are not challenging, fun or easily relatable. This is in line with research placing the average age of crowd workers at around 20-30 years \cite{kobayashi2015motivating} \cite{inproceedingsCHIAge2010}. On the other hand, crowd-volunteering tasks, often called citizen science tasks, such as the ones found on the Zooniverse platform \cite{zoonigeneralintro} can appeal to a more balanced representation of contributors, as about 15\% of the platform contributors self-report as retired.\footnote{Survey results were presented in a post: https://blog.zooniverse.org/2015/03/05/who-are-the-zooniverse-community-we-asked-them/}  There are also some crowdsourcing systems designed specifically for older adults which mitigate technology barriers, as in Hettiachchi et al. \cite{crowdtaskerchi2020} and tap into their knowledge and skills, such as tagging historical photos as in Yu et al. \cite{yu2016productive}, proofreading, as in Itoko et al. \cite{itoko2014involving} and Kobayashi et al. \cite{kobayashi2013age}, or both as in Skorupska et al. \cite{skorupska2019smartTV} They often rely on motivations that are pro-social, as in Kobayashi et al. \cite{kobayashi2015motivating} and also social, as in Seong et al. \cite{crowdolder2020chi} which is a trademark of Zooniverse. The Zooniverse platform allows crowd workers to support science projects at a larger scale by solving difficult tasks thanks to the impressive potential of such contributions \cite{zooniimpressivecrowdpotential} on a diverse crowdsourcing landscape of Zooniverse (www.zooniverse.org), which is why we have chosen this platform a to serve as the basis for this research. So, there is an opportunity to tap into the potential of older adults as crowd workers with a lot to offer and time on their hands - especially that their share in the society is increasing, and in 2019, "more than one fifth of the EU-27 population was aged 65 and over". \cite{population_structure2020} 

The question whether crowdsourcing tasks are effective in keeping older adults cognitively engaged is relevant, as volunteering activities \cite{morrow2010volunteering} in general may increase older adults' well-being \cite{morrow2003effects}, improve their mental and physical health \cite{lum2005effects} and can be seen as a protective factor for their psychological well-being \cite{greenfield2004formal,hao2008productive}, potentially delaying the onset of age-related issues \cite{kotteritzsch2014adaptive}. Therefore, in this study we want to gain insights into older adults' motivation and engagement with online citizen science tasks and uncover some guidelines for designing and presenting crowdsourcing citizen science tasks to this group. In designing our research we took care to uniformly present the wide-range of real crowd-volunteering tasks often appearing in citizen science projects. Only after older adults have completed each task we asked them how to improve it, and finally what would motivate them to engage with such tasks in the future.

\section{Methods}

In this study 33 older adults were asked to complete and evaluate 8 diverse, but standardized citizen science tasks at home, in an unsupervised environment. The study consisted of a short socio-demographic survey including questions about the participants' age, sex, education, activity, ICT-use, and crowdsorucing preferences based on Seong et al. \cite{crowdolder2020chi}. These questions were followed by a set of 8 different tasks chosen based on expert knowledge of the research team, localized into Polish and presented in an uniform way, broken into 4 pages - each page for a different type of a task. There were two tasks (one easier, and one more difficult/abstract) in each category of \textbf{image recognition (PIC)} for tasks T1 and T2, \textbf{audio recognition (AUD)} for tasks T3 and T4, \textbf{document transcription (DOC)} for tasks T5 and T6 and \textbf{pattern recognition (PAT)} for tasks T7 and T8, visible in Fig. \ref{visualoverview} in order. The tasks were selected out of 40 community-chosen projects active on the Zooniverse platform in the 2019-20 academic year and spotlit in the publication "Into the Zooniverse Vol. II",\footnote{The book is available for download here: https://blog.zooniverse.org/2020/11/17/into-the-zooniverse-vol-ii-now-available/} published on the 17th of November 2020. 

The final standardized tasks were as follows:

\begin{figure}[h]
  \centering
  \includegraphics[width=\linewidth]{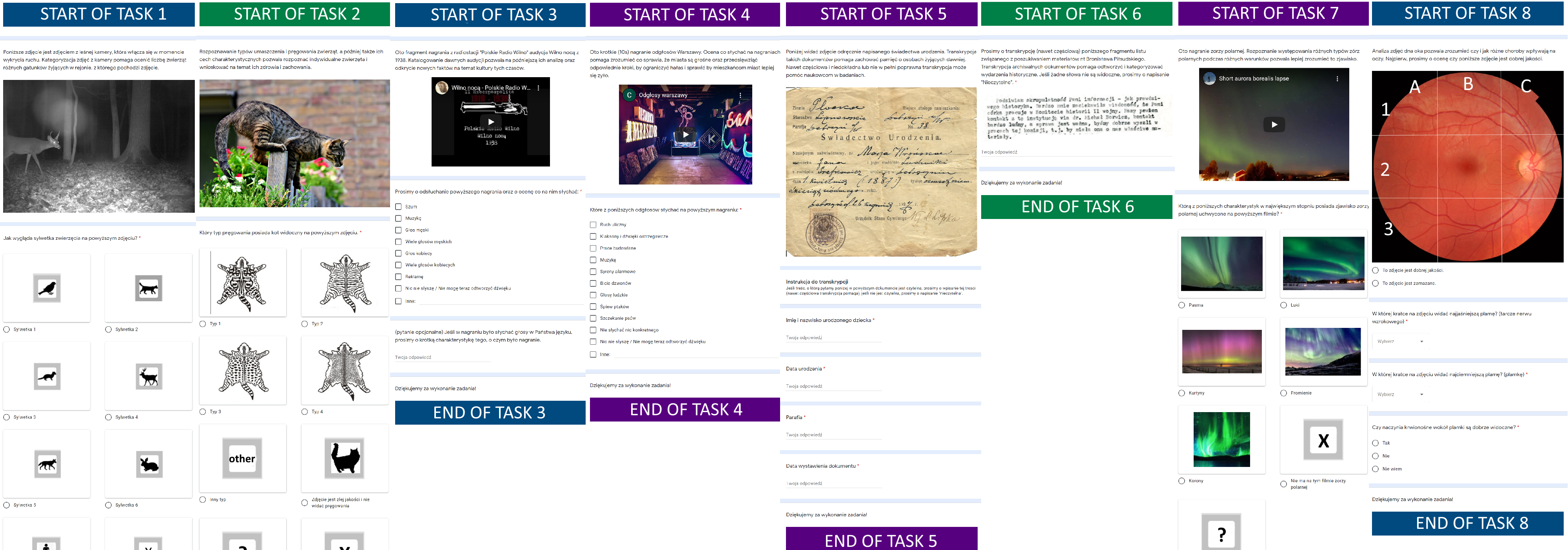}
  \caption{Visual overview of the tasks; T1-T8 from the left to the right.}
  \label{visualoverview}
  \vspace{-6mm}
\end{figure}

\begin{itemize}
\item \textbf{T1} Recognizing animal sillouettes - multiple choice of animal silhouettes, including human visible, no animal and other (representing MichiganZoomIn) 
\item \textbf{T2} Recognizing cat fur types on cat images - multiple choice with  abstract images of a cat pelts with fur patterns (similar to image recognition tasks)
\item \textbf{T3} Recognizing radio programs (97s-long recording) - checkboxes and a follow-up open answer about specifics (representing Vintage Cuban Radio)
\item \textbf{T4} Recognizing local urban sounds (10s-long recording) - checkboxes with pre-defined answers and an other option (representing Sounds of NYC)
\item \textbf{T5} Transcribing key information from a hand-written birth certificate of a person born in 1887 - 4 open short answer questions about the dates, name, and the location (representing tasks such as Every Name Counts)
\item \textbf{T6} Transcribing a longer (346 characters) typewritten text on a specific subject - 1 open long answer question (representing tasks relying on longer transcription of typewritten documents)
\item \textbf{T7} Recognizing Aurora Borealis patterns (6s-long recording) - multiple choice question with names of patterns and colours (representing Aurora Zoo)
\item \textbf{T8} Recognizing eye elements in eye pictures on a coordinate grid: two drop-down questions about coordinates and one multiple choice on the visibility of veins. (based on Eye for Diabetes; image by Mikael Häggström \cite{hmedical2014})
\end{itemize}
There was a short standardized introduction to each task explaining its importance and purpose, as not to bias the participants with the quality of the project presentation, which can vary considerably between projects. Then, the participants performed each example microtask. After each task we asked the participants to judge, on a 3-point scale, its:
\textbf{attractiveness}, \textbf{importance}, \textbf{ease of performing it}, \textbf{engagement}, and \textbf{if they would like to perform similar tasks in the future} and to suggest ways in which each task could be improved. The study protocol was positively evaluated by our ethics committee. The study itself was built in Google Forms and it took between 20-35 minutes to complete, depending on the amount of feedback given after each task and the ICT proficiency of the participants. Finally, after completing all tasks the participants were asked what could encourage them to engage in such tasks in general and whether they have done similar tasks in the past. The suggestions of motivating factors were inspired by an article by Campo et al \cite{campo_community_2019} as well as a wide body of research on crowdsourcing and volunteering.

\section{Results and Discussion}

\subsection{Participants}

There were 33 participants who completed and evaluated the chosen crowdsourcing tasks between Dec. 2020 and Feb. 2021. They were recruited from among the participants of our Living Lab \cite{kopec2017living} via e-mail as unpaid volunteers as we did not want to interfere with their motivation with a financial incentive. 22 participants were in the 60-69 age group, and 10 in the 70-79 group and 1 in 80+ group.\footnote{We have chosen to use multiple choice for age groups as not to bias the participants with an assumption that the research was targeted at older adults.} All of them were based in Poland, Polish and all but 6 of them came from larger cities (over 200k) and 21 of the participants had higher education. In Table \ref{tab:motipre} we can see the results concerning volunteering motivation before performing tasks for our 33 participants contrasted with results by Seong et al \cite{crowdolder2020chi}.

\begin{table}[h]
\centering
\scriptsize
\begin{tabular}{p{5cm}|p{1.8cm}|p{1.8cm}|p{1.5cm}|p{1.5cm}|}

 & \multicolumn{2}{p{3.6cm}|}{What would encourage you to engage with online or offline volunteer projects? n=33} & \multicolumn{2}{p{3cm}|}{Values older adults wanted from game experience \cite{crowdolder2020chi} n=12} \\ \hline
 & No. of P. & \% of P. & No. of P. & \% of P. \\ \hline
Physical improvement & 10 & 30.3\% & 3 & 25.0\% \\ \hline
Cognitive improvement & 15 & 45.5\% & 4 & 33.3\% \\ \hline
Opportunity to learn something new & 26 & 78.8\% & 5 & 41.7\% \\ \hline
Opportunity to communicate and interact with people & 14 & 42.4\% & 8 & 66.7\% \\ \hline
Opportunity to participate and contribute to society & 11 & 33.3\% & 4 & 33.3\% \\ \hline
None of the above & 4 & 12.1\% & - & - \\ \hline
\end{tabular}
\caption{Motivations of volunteer participants before the volunteer experience.}
\label{tab:motipre}
\vspace{-6mm}
\end{table}

Our participants use the following devices: 28 use a smartphone, 25 a laptop, 18 a desktop PC, 13 a tablet, 8 a SmartTV while 4 a smartwatch or a smartband and 2 a VR headset. They are also avid Internet users as 28 of them use the Internet either a few times a day or every day, and only 5 a few times a week or less often. As such, our participant group would be a good target for online volunteering and crowdsourcing tasks. Yet, after having completed the study 28 participants reported that they have never done similar tasks before, 3 of them said they did similar tasks at work and 2 did such tasks while volunteering.

\subsection{Performance and Feedback} 

\subsubsection{Image Recognition Tasks}
In \textbf{T1} 26 participants correctly identified the animal silhouette, 4 pointed to other silhouettes, 1 answered that there was no animal present while 2 more chose the "other" option where they have given in one case each: the name of the animal, the more detailed description of the animal. After completion 2 participants suggested to have a video instead - and 2 others wished the task was more challenging, while 1 complained the question was imprecise. Additionally, 1 person wished there were more animals to spot and "better hidden". In \textbf{T2} there was less agreement with 13 people choosing pattern 3, and 7 each voting for patterns 1 and 2, 2 for pattern 4, 3 saying it is "hard to tell" and 1 deciding it was some "other" pattern. The suggestions were to have "a different view of the cat in the picture" (1), a "couple of different pictures" (2) and comments appeared that "if someone does not like cats nothing can improve this task", but also "I liked it, even more so, because I like cats".

\subsubsection{Audio Recognition Tasks}
In both \textbf{T3} and \textbf{T4} our participants had no trouble listening to the recordings. In both tasks the majority of participants successfully identified the key audio elements (in T3: "many male voices" (31) in T4: "bells" (28) and "traffic" (25)). In T4 about half identified other elements ("birds singing", "people talking", "music"), while only one person noticed "barking", and one indicated "there was nothing specific" in the recording. Additionally, over half of the participants (18) chose to provide additional comment about the exact content of the radio recording from \textbf{T3}. For two people the \textbf{T3} recording was too short, for another too long and one wished it was accompanied by visuals. The feedback for \textbf{T4} was to have a longer recording (5), 1 person also wished for "more variety, to make it more difficult, but also more interesting" and to show a visual connected to the sound. One participant admitted that they "heard birds singing only upon the second hearing" but they were not sure. 

\subsubsection{Document Transcription Tasks}
Both transcription tasks were done very well. In \textbf{T5} only one participant decided that the text of the birth certificate was "intelligible" and only in 1 out of 4 places, while two people provided only the first name of the person. Among the others there is almost perfect agreement about what the text says with a varying level of detail for the name of the place and date notation. Additionally, one person provided a full transcription of the document, even though the task did not require it. Appearing suggestions were to have more information about the person in the birth certificate (2), to have more similar documents (1) and the participant suggesting it said that they have "transcribed 247 pieces of disappearing poetry before" and are experienced and now working on transcribing "very difficult historical letters". Another participant suggested to have documents related to the participants' own personal history (1). In \textbf{T6} most people (22) provided a complete transcription of the 346-character long text, on top of the transcription one person commented "(placing this dot here is incorrect - transcriber's comment)", while 3 wrote that the text was legible, and 6 provided an incomplete transcription, of these 2 added that the text was legible. Additionally, 2 wrote that it was intelligible. Two people wished for a more challenging text with a harder to read font, and one said other types of content interest them, other appearing comment was "For those who have not been involved in reading old manuscripts and other documents, this is a remarkably interesting activity (...) engaging and motivating, others will put it off or give up. I like it, it draws you in" while another participant mentioned that "such tasks require patience, they are not for everybody".

\subsubsection{Pattern Recognition Tasks}
In \textbf{T7} the count of choices was 12, 7, 6, 3, 3 for the dominant aurora pattern, while one person said that it is "hard to tell" and one person saw no aurora in the video. When asked about the colour 30 people agreed it was green, and over half added other colours (yellow, violet blue and pink). Here four people suggested to have a longer video, especially that "one would like to look longer, as we don't have that here and it is very interesting". In \textbf{T8} all participants (33) correctly identified the section coordinates with the described features; 26 said that the veins are clearly visible, 7 claiming to the contrary. One participant suggested that a longer analysis would improve this task, one more expressed that they are not sure where the macula of the retina was, and another wished for an analysis of some other organ. 

\subsubsection{Summary}
Overall, the older adults from our study in most cases provided high quality contributions with no training. Only T2 and T7 proved to be somewhat challenging and with these participants asked for more data.  Many wished for other tasks to be more challenging (harder font (T6), more audio variety (T4), more and better hidden animals (T1), longer analysis (T8)) and the "easy" dimension had the weakest correlations with willingness to do similar tasks in the future. It seems that older adults would not mind, and even preferred it, if the tasks posed more of a challenge (eg. T5 vs T6), especially if it would allow them to learn something interesting. They also wished for the shorter tasks to be extended, either by additional data (T1, T2) steps (T8) or longer duration (T7, T3, T4), not only because they enjoyed them and wanted to learn more, but also to allow them to provide higher quality contributions by adding more data to verify their choices. It seems therefore, that microtasks, designed to be brief for efficiency, could be extended and elaborated upon to increase the contributors' satisfaction, especially if they rely on image, video or audio data. 

\subsection{Evaluation of Tasks}

Participants rated T8 the highest, while T2 the lowest. They distributed most points in the category "I would do similar tasks" (380) followed by "easy" (370), "attractive" (363), "engaging" (328) and "important" (283). Our participants, once exposed to each task, reported a high willingness to engage with similar tasks (with an average of 1.44 out of 2) suggesting, that older adults would engage with such tasks more, if they were made more easily available to them.

\begin{figure}[ht]
  \centering
  \includegraphics[width=\linewidth]{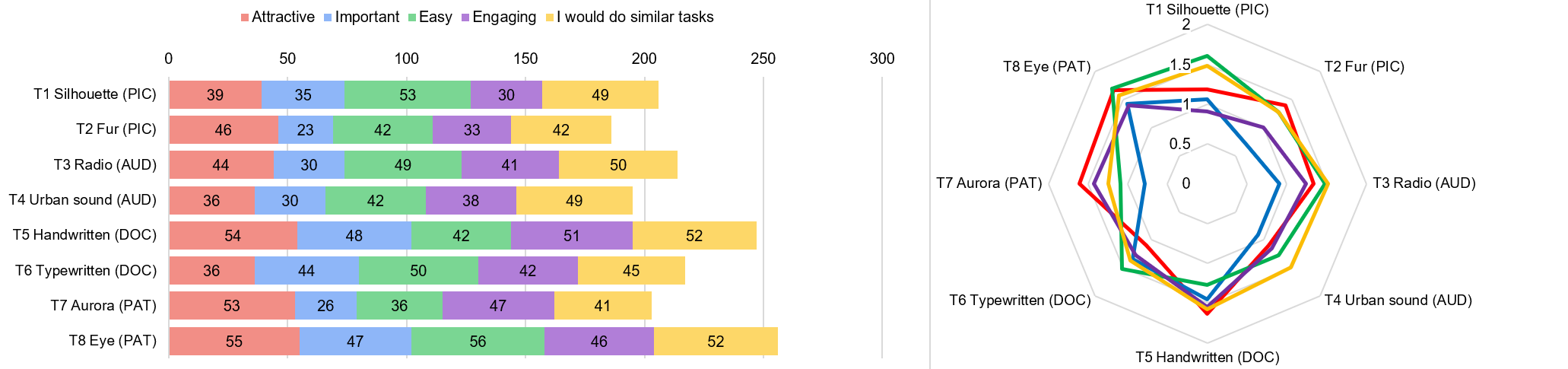}
  \caption{Left: Total points awarded by our participants after the completion of each task. Right: Average scores for the same tasks.}
  \label{totalpoints}
\end{figure}

\begin{figure}[ht]
  \centering
  \includegraphics[width=\linewidth]{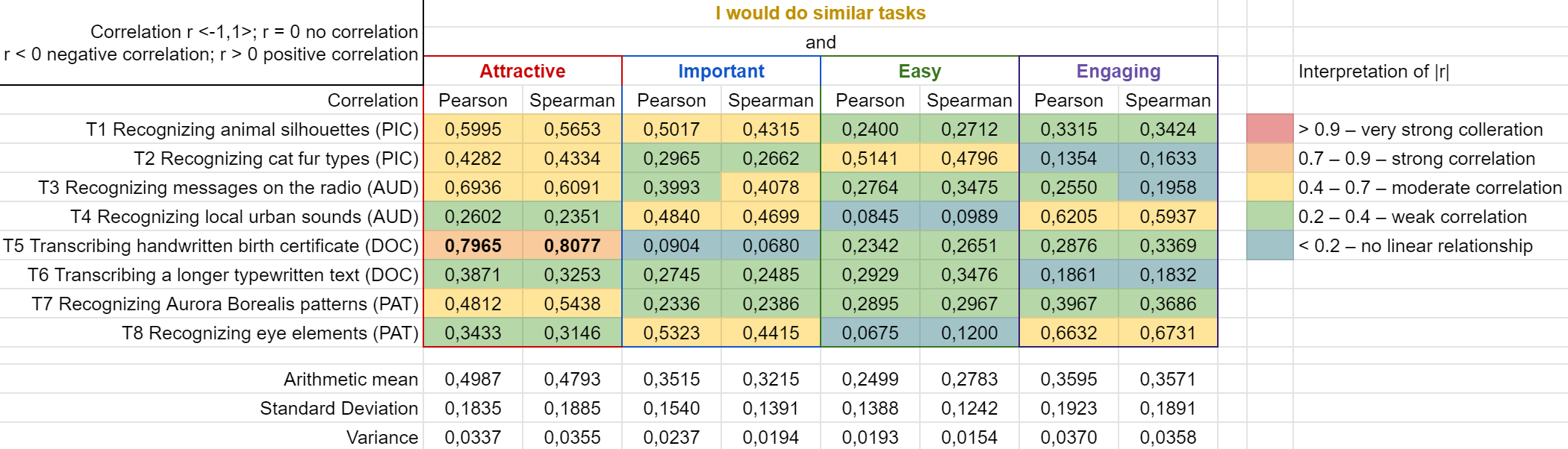}
  \caption{Correlation matrix for the dimension "I would do similar tasks" and other dimensions, from older adults' evaluation of the tasks right after performing them.}
  \label{matrix}
  \vspace{-8mm}
\end{figure}

As seen in Fig. \ref{matrix} the correlations with the willingness to do similar tasks in the future are either positive, or close to zero, while the strongest correlation is with the visual or thematic "attractiveness" of the task. It was also slightly important whether the task was "engaging" or "important", especially if it was not found to be "attractive" and to a lesser extent if it was "easy". This suggests that older adults' main motivation is rather intrinsic, connected to their own interest in the task, which of course is moderated by other variables.

\subsection{Motivation}

After having completed all of the tasks, the participants chose learning something new and information about the purpose of performing these tasks as the prevailing motivators. They would also like to receive feedback on their performance and to have detailed tutorials. Detailed results are reported in Table \ref{tab:motipost}. Moreover, the ability to perform these tasks using interfaces (smartphone, Smart TV or audio) other than a computer screen was judged as not particularly important, however, this may be due to lack of familiarity with them for audio and TV devices, as there are studies which successfully implement them for crowdsourcing \cite{crowdtaskerchi2020,skorupska2018smarttv} and the challenge of small-screen interaction for smartphones \cite{olderscreeninter2011} which is still relevant. \cite{kowalski2019voice}

\begin{table}[!htbp]
\scriptsize
\begin{tabular}{l|l|l}
                                                                               & No. of P. & \% of P. \\ \hline
The   opportunity to learn something interesting while performing these tasks  & 24           & 72.7\%      \\ \hline
More knowledge about the     purpose of performing these tasks                 & 24           & 72.7\%      \\ \hline
Receiving feedback on the     use and usefulness of the tasks performed        & 21           & 63.6\%      \\ \hline
A short training to make     sure I do the tasks well                          & 17           & 51.5\%      \\ \hline
More interesting topic of     tasks                                            & 9            & 27.3\%      \\ \hline
Online support and contact     with other people performing these tasks        & 9            & 27.3\%      \\ \hline
Tasks suited to my skills                                                      & 9            & 27.3\%      \\ \hline
Training and personal    meetings for those performing the tasks               & 7            & 21.2\%      \\ \hline
Statistics showing the number    of already completed tasks                    & 7            & 21.2\%      \\ \hline
The ability to perform these    tasks on the TV screen with the remote control & 6            & 18.2\%      \\ \hline
Thanks from the researchers                                                    & 6            & 18.2\%      \\ \hline
Ability to perform these    tasks on a smartphone                              & 5            & 15.1\%      \\ \hline
Ability to perform these    tasks using the voice interface                    & 0            & 0.0\%       \\ \hline
"None of the above" and "Other (own answer)"                                                            & 0            & 0.0\%  \\ \hline  
\end{tabular}
\caption{Answers to "Which of these elements would encourage you to perform tasks similar to the sample tasks in this survey?" asked after completing all of the tasks.}
\label{tab:motipost}
\vspace{-6mm}
\end{table}

\section{Conclusions}

In this exploratory research we have verified that crowdsourcing microtasks, especially those appearing in citizen science projects, can be well-suited for some groups of older adults - both in terms of the quality of older adults' contributions and their motivation. Yet, even among older adults with average and higher ICT skills - sufficient to contribute to such projects, such as the participants in our sample, the awareness of the existence of such crowdsourcing projects is quite low, as such citizen science tasks are not easily found and sampled. Older adults as a group often overlooked as potential contributors to larger scale crowdsourcing projects due to their often lower willigness to engage online and the perception of their ICT skills. However, the older adults in our study who received no compensation, provided high quality contributions with little training and were open to continue volunteering online. 

To increase participation, and thus the representation of this age group's voice in citizen science, we suggest that crowdsourcing tasks ought to be advertised in line with older adults' preferences. These are related to the way in which completing these tasks may benefit, first, them individually, and then, the society as a whole. Based on our research, crowdsourcing microtasks' presentation should focus on the aspect of learning something interesting (which was confirmed by an arithmetic mean correlation of 0.47 for "I would do similar tasks" and "Attractive, thematically or visually"), rather than the aspect of being able to utilize ones' existing skills and knowledge. The contributors should also be provided with a high awareness of the tasks' purpose and ought to be made aware of the usefulness of their individual contributions to reassure the participants that it was time well spent. The tasks could also be more elaborate, to provide an appropriate challenge and increase immersion. Hence, in future research we would also like to examine a wider range of tasks of increasing complexity and duration, as well as the effects of engaging in crowdsourcing on participant's physical, mental or cognitive well-being in further comparative longitudinal research with larger groups of participants of all ages.  

\bibliographystyle{splncs04}
\bibliography{mybibliography}

\begin{thebibliography}{10}
\providecommand{\url}[1]{\texttt{#1}}
\providecommand{\urlprefix}{URL }
\providecommand{\doi}[1]{https://doi.org/#1}

\bibitem{aula_learning_2004}
Aula, A.: Learning to use computers at a later age. In: {HCI} and the {Older}
  {Population}, pp.~3--5. University of Glasgow, Leeds, UK (Sep 2004)

\bibitem{zooniimpressivecrowdpotential}
Barber, S.T.: The zooniverse is expanding: Crowdsourced solutions to the hidden
  collections problem and the rise of the revolutionary cataloging interface.
  Journal of Library Metadata  \textbf{18}(2),  85--111 (2018).
  \doi{10.1080/19386389.2018.1489449}

\bibitem{brewer2016would}
Brewer, R., Morris, M.R., Piper, A.M.: Why would anybody do this?:
  Understanding older adults' motivations and challenges in crowd work. In:
  Proceedings of the 2016 CHI Conference on Human Factors in Computing Systems.
  pp. 2246--2257. ACM (2016)

\bibitem{campo_community_2019}
Campo, S.a., Khan, V.J., Papangelis, K., Markopoulos, P.: Community heuristics
  for user interface evaluation of crowdsourcing platforms. Future Generation
  Computer Systems  \textbf{95},  775 -- 789 (2019)

\bibitem{djoub_ict_2013}
Djoub, Z.: {ICT} education and motivating elderly people. In: Ariadna; cultura,
  educación y tecnología. vol.~1, pp. 88--92 (Jul 2013)

\bibitem{population_structure2020}
Population structure and ageing,
  \url{https://ec.europa.eu/eurostat/statistics-explained/index.php}

\bibitem{greenfield2004formal}
Greenfield, E.A., Marks, N.F.: Formal volunteering as a protective factor for
  older adults' psychological well-being. The Journals of Gerontology Series B:
  Psychological Sciences and Social Sciences  \textbf{59}(5),  S258--S264
  (2004)

\bibitem{hao2008productive}
Hao, Y.: Productive activities and psychological well-being among older adults.
  The Journals of Gerontology Series B: Psychological Sciences and Social
  Sciences  \textbf{63}(2),  S64--S72 (2008)

\bibitem{von2018influence}
von Helversen, B., Abramczuk, K., Kope{\'c}, W., Nielek, R.: Influence of
  consumer reviews on online purchasing decisions in older and younger adults.
  Decision Support Systems  (2018)

\bibitem{crowdtaskerchi2020}
Hettiachchi, D., Sarsenbayeva, Z., Allison, F., van Berkel, N., Dingler, T.,
  Marini, G., Kostakos, V., Goncalves, J.: "hi! i am the crowd tasker"
  crowdsourcing through digital voice assistants. In: Proceedings of the 2020
  CHI Conference on Human Factors in Computing Systems. p. 1–14. CHI '20,
  Association for Computing Machinery, New York, NY, USA (2020).
  \doi{10.1145/3313831.3376320}

\bibitem{hmedical2014}
Häggström, M.: Medical gallery of mikael häggström 2014 (Jul 2014),
  \url{https://en.wikiversity.org/}

\bibitem{itoko2014involving}
Itoko, T., Arita, S., Kobayashi, M., Takagi, H.: Involving senior workers in
  crowdsourced proofreading. In: International Conference on Universal Access
  in Human-Computer Interaction. pp. 106--117. Springer (2014)

\bibitem{knowles2018older}
Knowles, B., Hanson, V.L.: Older adults' deployment of 'distrust'. ACM Trans.
  Comput.-Hum. Interact.  \textbf{25}(4),  21:1--21:25 (Aug 2018).
  \doi{10.1145/3196490}, \url{http://doi.acm.org/10.1145/3196490}

\bibitem{knowles2018wisdom}
Knowles, B., Hanson, V.L.: The wisdom of older technology (non)users. Commun.
  ACM  \textbf{61}(3),  72--77 (Feb 2018). \doi{10.1145/3179995},
  \url{http://doi.acm.org/10.1145/3179995}

\bibitem{kobayashi2015motivating}
Kobayashi, M., Arita, S., Itoko, T., Saito, S., Takagi, H.: Motivating
  multi-generational crowd workers in social-purpose work. In: Proceedings of
  the 18th ACM Conference on Computer Supported Cooperative Work \& Social
  Computing. pp. 1813--1824. ACM (2015)

\bibitem{olderscreeninter2011}
Kobayashi, M., Hiyama, A., Miura, T., Asakawa, C., Hirose, M., Ifukube, T.:
  Elderly user evaluation of mobile touchscreen interactions. In: Proceedings
  of the 13th IFIP TC 13 International Conference on Human-Computer Interaction
  - Volume Part I. p. 83–99. INTERACT'11, Springer-Verlag, Berlin, Heidelberg
  (2011)

\bibitem{kobayashi2013age}
Kobayashi, M., Ishihara, T., Itoko, T., Takagi, H., Asakawa, C.: Age-based task
  specialization for crowdsourced proofreading. In: International Conference on
  Universal Access in Human-Computer Interaction. pp. 104--112. Springer (2013)

\bibitem{kopec2017living}
Kope\'{c}, W., Skorupska, K., Jaskulska, A., Abramczuk, K., Nielek, R.,
  Wierzbicki, A.: Livinglab pjait: Towards better urban participation of
  seniors. In: Proceedings of the International Conference on Web Intelligence.
  pp. 1085--1092. WI '17, ACM, New York, NY, USA (2017).
  \doi{10.1145/3106426.3109040}

\bibitem{kotteritzsch2014adaptive}
K{\"o}tteritzsch, A., Koch, M., Lem{\^a}n, F.: Adaptive training for older
  adults based on dynamic diagnosis of mild cognitive impairments and dementia.
  In: International Workshop on Ambient Assisted Living. pp. 364--368. Springer
  (2014)

\bibitem{kowalski2019voice}
Kowalski, J., Jaskulska, A., Skorupska, K., Abramczuk, K., Biele, C.,
  Kope{\'c}, W., Marasek, K.: Older adults and voice interaction: A pilot study
  with google home. In: Extended Abstracts of the 2019 CHI Conference on Human
  Factors in Computing Systems. pp. 187:1--187:6. CHI EA '19, ACM, New York,
  NY, USA (2019). \doi{10.1145/3290607.3312973}

\bibitem{lum2005effects}
Lum, T.Y., Lightfoot, E.: The effects of volunteering on the physical and
  mental health of older people. Research on aging  \textbf{27}(1),  31--55
  (2005)

\bibitem{morrow2010volunteering}
Morrow-Howell, N.: Volunteering in later life: Research frontiers. The Journals
  of Gerontology Series B: Psychological Sciences and Social Sciences
  \textbf{65}(4),  461--469 (2010)

\bibitem{morrow2003effects}
Morrow-Howell, N., Hinterlong, J., Rozario, P.A., Tang, F.: Effects of
  volunteering on the well-being of older adults. The Journals of Gerontology
  Series B: Psychological Sciences and Social Sciences  \textbf{58}(3),
  S137--S145 (2003)

\bibitem{inproceedingsCHIAge2010}
Ross, J., Irani, L., Silberman, M.S., Zaldivar, A., Tomlinson, B.: Who are the
  crowdworkers? shifting demographics in mechanical turk. In: CHI '10 Extended
  Abstracts on Human Factors in Computing Systems. p. 2863–2872. CHI EA '10,
  Association for Computing Machinery, New York, NY, USA (2010).
  \doi{10.1145/1753846.1753873}

\bibitem{crowdolder2020chi}
Seong, E., Kim, S.: Designing a crowdsourcing system for the elderly: A
  gamified approach to speech collection. In: Extended Abstracts of the 2020
  CHI Conference on Human Factors in Computing Systems. p. 1–9. CHI EA '20,
  Association for Computing Machinery, New York, NY, USA (2020).
  \doi{10.1145/3334480.3382999}

\bibitem{zoonigeneralintro}
Simpson, R., Page, K.R., De~Roure, D.: Zooniverse: Observing the world's
  largest citizen science platform. In: Proceedings of the 23rd International
  Conference on World Wide Web. p. 1049–1054. WWW '14 Companion, Association
  for Computing Machinery, New York, NY, USA (2014).
  \doi{10.1145/2567948.2579215}

\bibitem{skorupska2018smarttv}
Skorupska, K., N\'{u}\~{n}ez, M., Kope{\'c}, W., Nielek, R.: Older adults and
  crowdsourcing: Android tv app for evaluating tedx subtitle quality. Proc. ACM
  Hum.-Comput. Interact.  \textbf{2}(CSCW),  159:1--159:23 (Nov 2018).
  \doi{10.1145/3274428}

\bibitem{skorupska2019smartTV}
Skorupska, K., N{\'u}{\~{n}}ez, M., Kope{\'{c}}, W., Nielek, R.: A comparative
  study of younger and older adults' interaction with a crowdsourcing android
  tv app for detecting errors in tedx video subtitles. In: Lamas, D., Loizides,
  F., Nacke, L., Petrie, H., Winckler, M., Zaphiris, P. (eds.) Human-Computer
  Interaction -- INTERACT 2019. pp. 455--464. Springer International
  Publishing, Cham (2019)

\bibitem{yu2016productive}
Yu, H., Miao, C., Liu, S., Pan, Z., Khalid, N.S.B., Shen, Z., Leung, C.:
  Productive aging through intelligent personalized crowdsourcing. In: 30th
  AAAI Conference on Artificial Intelligence (AAAI-16) (2016)

\end{thebibliography}

\end{document}